\documentclass[prl,twocolumn,amsmath,showpacs,superscriptaddress,floatfix]{revtex4}
\usepackage{graphicx}

\begin{document}
\title{Force Dynamics in Weakly Vibrated Granular Packings}

\author{Paul Umbanhowar}

\affiliation{Department of Physics and Astronomy, Northwestern
University, Evanston, IL 60208, USA.}

\author{Martin van Hecke}

\affiliation{Kamerlingh Onnes Lab, Leiden University, PO box 9504,
2300 RA Leiden, The Netherlands.}
\date{\today}

\begin{abstract}
The oscillatory force $F_b^{ac}$ on the bottom of a rigid,
vertically vibrated, grain filled column, reveals rich granular
dynamics, even when the peak acceleration of the vibrations is
significantly less than the gravitational acceleration at the
earth's surface. For loose packings or high frequencies,
$F^{ac}_b$'s dynamics are dominated by grain motion. For moderate
driving conditions in more compact samples, grain motion is
virtually absent, but $F^{ac}_b$ nevertheless exhibits strongly
nonlinear and hysteretic behavior, evidencing a granular regime
dominated by nontrivial force-network dynamics.
\end{abstract}

\maketitle

Granular media consist of macroscopic solid grains which interact
via dissipative, repulsive contact forces. Thermal energy is
inconsequential, and granulates {\em jam} in random configurations
unless sufficient mechanical energy is supplied, for example by
shearing or shaking \cite{jamming_gospel}. For sinusoidally,
vertically vibrated granular media the driving strength is
characterized by the nondimensional acceleration amplitude, $\Gamma
\!=\! A (2 \pi f)^2/g$, where $A$ is the displacement amplitude, $f$
is the oscillation frequency, and $g$ is the gravitational
acceleration at the earth's surface. In the well-studied case of
$\Gamma \agt 1$, grains periodically lose contact with and
subsequently impact the oscillating container. The collision, as
well as the accompanying relative motion between grains and wall,
injects energy into the system that drives grain rearrangements
\cite{bougie,nmat}.

In this Letter we explore the $\Gamma < 1$ regime by measuring the
oscillatory force on the container bottom, $F_b^{ac}$, in a weakly
vibrated column filled with grains of total mass $M$ [Fig.~1(a)].
%
%
$F_b^{ac}$ exhibits rich dynamics. Under two conditions the dynamics
is dominated by grain rearrangements: {\em{(i)}} Loose packings may
compact when they are first subjected to weak vibrations, which
leads to intermittent burst in $F_b^{ac}$. {\em{(ii)}} For large
driving frequencies, grains slide periodically with respect to the
container even at low $\Gamma$ (e.g.~$\Gamma\approx 0.1$ at
$f=900$~Hz), which leads to strongly nonlinear behavior in
$F_b^{ac}$.

With moderate driving in compacted samples, however, {\em relative
grain motion is minute but variations in the force configuration
remain substantial} as indicated by nontrivial changes in
$F_b^{ac}$. Weak vibrations thus excite strongly nonlinear and
glassy dynamics of the force network. Such force variations are
allowed, since for a terrestrial granulate the grain and deformation
scales are separated by many orders of magnitude \cite{jacco}: a
700~$\mu$m diameter bronze sphere is compressed only $\sim\!100$ nm
under the weight of 1000 additional identical spheres
\cite{kljohnson}.
%

\begin{figure}[t]
\includegraphics[width=\columnwidth]{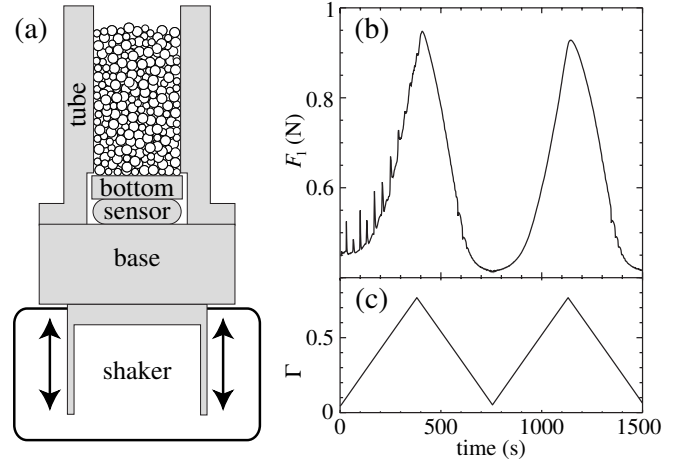}
\caption[]{(a) Schematic of the experiment showing the piezo force
sensor mounted between the ``bottom" and the ``base" (diameter
$\times$ height: $32\times 12$~mm and $ 89\times 62$~mm
respectively), with a cylindrical tube (inner/outer diameter
30/55~mm, height 113~mm) which is attached only to the base. The
small $\sim\!100$~$\mu$m gap between the tube and bottom plate
prevents grains from becoming trapped. Shaded parts are rigidly
connected and move in unison. The sensor signal is used to obtain
$F_b^{ac}$, the vertical AC force exerted on the bottom plate by the
grains. (b,c) Nonlinear response of a column filled with 200~g of
$0.61-0.70$~mm diameter bronze particles under sweeps of the
vibration amplitude $\Gamma$ for $f\!=\!80$~Hz; $F_1$ denotes the
calibrated ratio of the first harmonic of $F_b^{ac}$ to $\Gamma$
(Eq.~\ref{def_f}). } \label{figbasic} \vspace{-4.1mm}
\end{figure}

Figures~\ref{figbasic}(b,c) illustrate the main features of $F_1$,
the first harmonic of $F_b^{ac}$ divided by $\Gamma$ [see
Eq.~(\ref{def_f})], for a full column of grains undergoing slow
triangular sweeps of $\Gamma$. For a solid mass placed on the bottom
plate, $F_1$ is independent of $\Gamma.$ For grains, however, $F_1$
depends strongly and nonlinearly on $\Gamma$ [e.g.~$F_1(\Gamma=0.5)
\approx 2 F_1(\Gamma =0.05)$], even for low driving frequencies. The
spikes in $F_1$ during the initial ramp are caused by compaction of
the material. The asymmetry of $F_1(\Gamma)$ indicates hysteresis
and memory in compacted samples. We stress that the strength of
these features does not vary significantly for driving frequencies
from 16 to 300~Hz. The phenomena evident in Fig.~\ref{figbasic}(b)
are essentially {\em quasistatic} and are not associated with the
excitation of sound waves (see below).

{\em Experimental Setup --- } Nearly-spherical bronze particles
sieved between 0.61 and 0.70~mm are poured into a smooth cylindrical
tube with a detached bottom which is supported by a rigid
piezo-electric force sensor (stiffness $2.5$~GN/m). The {\em entire
assembly} is vertically oscillated with a small sinusoidal
displacement [Fig.~1(a)]. An accelerometer attached to the tube
measures the time-resolved acceleration $\gamma(t)$ which, for most
driving conditions, is harmonic, equaling $ \Gamma \sin(2 \pi f t)$.
The measured force is sensitive to temperature drift. The entire
assembly is therefore placed in a temperature controlled enclosure
maintained slightly above room temperature (temperature fluctuations
$\pm$ 10 mK, humidity 5-10\%); grains are equilibrated in the
enclosure prior to use~\cite{in_preparation}.

The deflection of the relatively compliant force sensors used in
most previous studies is large compared to the deformation of hard
grains like steel or glass; the granular force configuration is then
completely altered due to relative motion between force probe and
grains. In contrast, our tube/sensor assembly is effectively a {\em
solid} container since the maximal deflection of the piezo is less
than 1~nm, which ensures that the measured force variations are
intrinsic to the granular
medium~\cite{eric,memories,anna_plus_comment}.

In our experiment, the force sensor measures the total {\em AC
force}, $F_{total}^{ac}$, which is the sum of the inertial force
generated by the acceleration of the bottom plate and sensor, with
effective mass $M_0,$ and the AC bottom force $F_b^{ac}$ resulting
from the acceleration of material in the column (which can have
harmonics):
\begin{equation}\label{def_f}
F_{total}^{ac} = \Gamma M_0 g  \sin(2 \pi f t) + \Gamma
\sum_{n=1}^\infty F_n \sin (2 n\pi ft-\phi_n)~.
\end{equation}
To calibrate the signal, $F_{total}^{ac}$ was measured for a range
of $f$ and $\Gamma$, both with and without solid test masses
attached to the bottom plate.  The value of $F_{total}^{ac}$ for the
empty system allows us to subtract the term $\propto \Gamma M_0 g $,
after which $F_1$ is found to be proportional to the test mass and
independent of $\Gamma$ (the definition of $F_n$ isolates the
trivial scaling with $\Gamma$). The higher harmonics $F_2,F_3,
\dots$ are negligible in this case.

We now consider the AC vertical components of the frictional
wall-force $F_w$ and the bottom force $F_b$ in a grain filled
column. We will distinguish a {\em contact regime} where grains do
not slide with respect to the column, and an {\em impact regime}
where they do. In the impact regime,
\begin{equation}
M g \Gamma \sin (2 \pi f t) = F_w^{ac} + F_b^{ac}~, \label{eqm3}
\end{equation}
and our measurements probe how the vibration induced variations of
the grain weight, $M g \Gamma \sin (2 \pi f t)$, are distributed
between wall and bottom. However, Eq.~(\ref{eqm3}) is violated in
the impact regime. We have checked the validity of Eq.~(\ref{eqm3})
by placing a separate, {\em closed bottom}, grain filled container
directly on the bottom plate \cite{newrefs}, so that the sensor
measures the sum $F_w^{ac} + F_b^{ac}$; when the sensor signal
remains purely harmonic, the system is in the contact regime, when
(strong) nonlinearities are detected, it is in the impact regime. As
we will show for the contact regime, $F_b^{ac}$ grows nonlinearly
with grain mass and $\Gamma$, exhibits hysteresis and memory
effects, and higher harmonics play a role---very different behavior
than for a solid mass. However, we first discuss grain motion
dominated {\em compaction} and {\em impact}.

\begin{figure}[tbp]
\includegraphics[width=0.8\columnwidth]{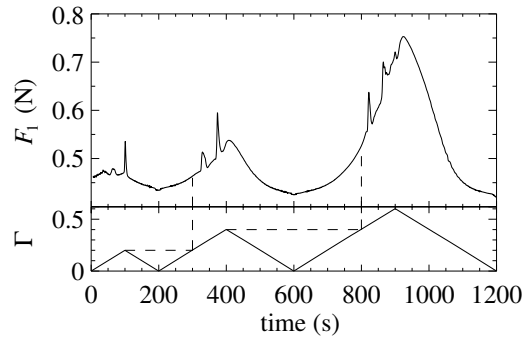}
\caption[]{Spikes observed for $f\!=\!80$~Hz and $M\!=\!200$~g in an
initially low density packing. Sweeps of increasing magnitude in
$\Gamma$ illustrate that spikes only occur in ``fresh'' territory.
}\label{fig_rearrangement} \vspace{-4.1mm}\end{figure}

{\em Compaction} --- ``Spikes'', such as those shown in
Fig.~\ref{figbasic}(b), occur when $\Gamma$ is ramped up in loosely
packed samples, which are formed by placing the end of a funnel on
the bottom of the container, filling the funnel with material, and
then slowly retracting the funnel. Figure~\ref{fig_rearrangement}
illustrates that spikes only occur in ``fresh'' territory, {\it
i.e.}, when $\Gamma$ is increased beyond its previous maximum value.
After $\Gamma$ has been swept up to a value near one, we refer to
the sample as {\em fully annealed}.

During a spike, which typically lasts for 1000's of oscillation
cycles, $\phi_1$ shifts significantly (indicating dissipation) and
$F_b^{ac}$ is strongly non-sinusoidal. The specific $\Gamma$ values
where spikes occur vary from run to run, and hence are not resonant
effects.  Spikes are apparently due to compaction, since the free
surface is lower after a spike occurs, and gently poured columns
with lower initial density produce more spikes than less gently
prepared ones with higher initial density. Spikes and their
dependence on $\Gamma$ apparently result from the full mobilization
of frictional forces at the walls in granular columns
\cite{janssen,eric}. Under these conditions, vertical frictional
forces are near their maximal values so that vibrations can cause
(micro) slippage, and stronger vibrations may cause further
compaction. The behavior of spikes is not frequency dependent:
qualitatively similar spikes occur for $f$ between 16 and 300~Hz,
and switching to a different frequency in a partially annealed
sample does not appreciably alter the $\Gamma$ value separating
fresh and annealed states \cite{in_preparation}.

\begin{figure}[btp]
\includegraphics[width=\columnwidth]{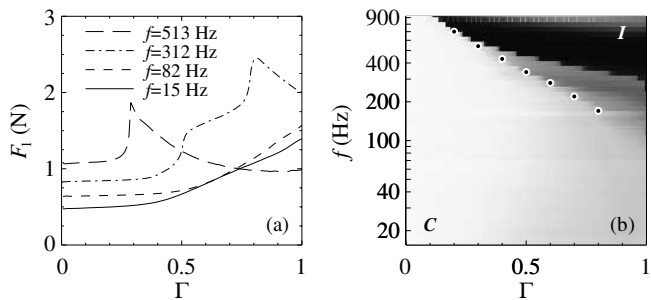}
\caption[]{Contact and impact regimes ($M=200$~g): (a)
$F_1(\Gamma,f)$ is smooth at low frequencies, but increases abruptly
and peaks for higher frequencies (data is vertically offset for
clarity). (b) Greyscale intensity plot of nonlinearity
${\cal{N}}(\Gamma,f) \!=\! F_1(\Gamma,f)- F_1(0,f)$ (white,
${\cal{N}} \!=\!0$; black, ${\cal{N}}$ positive) indicating the
contact ($C$) and impact ($I$) regimes. Dots indicate the initial
sharp increase in $F_1$. }\label{fig4} \vspace{-4.1mm}\end{figure}

{\em Impact vs Contact ---} For compacted samples we distinguish
between the contact regime where Eq.~\ref{eqm3} is satisfied, and
the impact regime where Eq.~\ref{eqm3} is violated. Figure
\ref{fig4}(a) illustrates that for low frequencies $F_1$ increases
smoothly with $\Gamma$, while for higher frequencies there is a
sudden upturn and a peak. In the vicinity of and above this
transition, $\gamma(t)$ and $F_b^{ac}$ are strongly anharmonic
--- as when grains periodically {\em collide} with the bottom for
$\Gamma >1$; this is the impact regime. Below the transition,
$\gamma(t)$ remains sinusoidal; this is the contact regime.

To check that Eq.~(\ref{eqm3}) is valid in the contact regime ({\em
i.e.}~no slipping) but not in the impact regime, we replaced the
open-bottom tube with a similarly sized, but {\em closed-bottom}
grain filled tube mounted directly on the bottom plate. In this
configuration, $F_b^{ac}$ is insensitive to smooth force transfers
between the wall and the container bottom which allows
Eq.~(\ref{eqm3}) to be checked directly. In the impact regime
$F_b^{ac}$ and $\gamma(t)$ are strongly anharmonic in violation of
Eq.~(\ref{eqm3}); in the contact regime they remain sinusoidal with
$F_1$ essentially independent of $\Gamma$ in agreement with
Eq.~(\ref{eqm3}). The physics underlying the force dynamics in the
contact regime is thus a smooth, periodic transfer of grain weight
between wall and bottom.


Figure~\ref{fig4}(b) displays the strength of the nonlinearity of
$F_1$ as function of $\Gamma$ and $f$---the rapid increase in
nonlinearity
marks the onset of impact. Note that at high frequency ($f\approx
1$~kHz), the impact regime occurs for surprisingly small $\Gamma$
($\approx\!0.1$). Apparently, impact for $\Gamma<1$ is due to the
excitation of resonant granular sound waves. Typical sound speeds
are of the order of 100~ms$^{-1}$ \cite{soundspeed}, so in our 10~cm
deep column we expect a resonant response around 1~kHz. This picture
is consistent with findings of Yanagida {\em et~al.}~\cite{newrefs}
in studies of the resonant response of grain filled closed bottom
containers for small $\Gamma$, and it is also consistent with the
shift of the impact transition to higher frequencies for smaller $M$
(this also excludes a trivial resonance of the
apparatus\cite{in_preparation}). The $\Gamma$ dependence of this
transition is not fully understood.
%

{\em Contact Regime Nonlinearity --- } We now explore the nonlinear response of
fully annealed samples in the contact regime as a function of $M$
and $\Gamma$. The frequency is fixed at $80$~Hz, since in the
contact regime the grain response varies only weakly with $f$.

\begin{figure}[tbp]
\includegraphics[width=1\columnwidth]{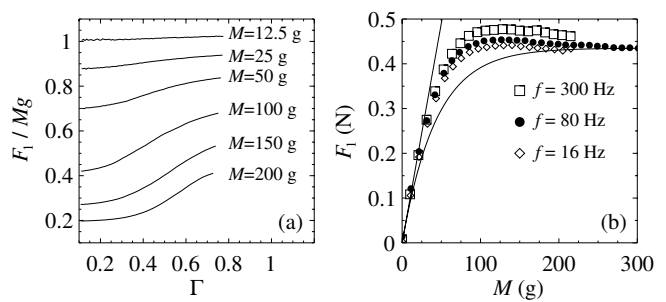}
\caption[]{Nonlinearity of $F_1$ with $\Gamma$ and total grain mass
$M$. (a) $F_1(\Gamma)/Mg$ for various filling fractions ($f\!=\!80$
Hz). For larger $M$ and fixed $\Gamma$, the weight fraction on the
bottom decreases. (b) $F_1$ as function of $M$ in the linear regime
($\Gamma\!=\!0.05$) for $f\!=\!$ 16, 80 and 300 Hz, compared to a
linear response (straight line) and a Janssen-like response
 $F_1/F_{sat} = 1 -
\exp(-M g / F_{sat})$ for $F_{sat}=0.435$~N.}\label{jansen}
\vspace{-4.1mm}\end{figure}

Figure~\ref{jansen}(a) illustrates that the small mass behavior is
independent of $\Gamma$, and that the grains are supported entirely
by the bottom since \mbox{$F_1 \approx M g$}. For larger masses,
wall forces start to play a role since \mbox{$F_1 < M g$} and the
material's response becomes increasingly nonlinear with $\Gamma$:
the exchange between bottom and wall forces underlies the
nonlinearity of $F_1(\Gamma)$ for large $M$. This strong
nonlinearity is accompanied by a modest increase in the harmonic
distortion \cite{thd} of $F_b^{ac}$ and a similarly modest shift in
$\phi_1$. For example for $\Gamma \!=\! 0.7$ and $M\!=\!200$~g,
$F_1$ is doubled from its low $\Gamma$ value, while the harmonic
distortion is only 6\% and $\phi_1 \approx 6^{\circ}$. Underlying
the harmonic distortion is the nonlinear (Hertzian) contact-force
law which makes grain contacts stiffer when under larger pressure
\cite{kljohnson}; the waveform distortions we have obtained for
$F_b^{ac}(t)$  are consistent with this picture
\cite{in_preparation}.

For small $\Gamma$ the response is linear in $\Gamma$: for
$\Gamma\!\approx\!0.05$, $F_b^{ac}$ is harmonic ($<1$\% distortion)
and in-phase with the acceleration, and $F_1$ varies less than $1$\%
for $0<\Gamma<0.1$. We study $F_1(M)$ as a function of increasing
mass $M$ at $\Gamma\!\approx\!0.05$ by incrementally pouring grains
from a height of approximately 10~cm above the grain surface. Figure
\ref{jansen}(b) illustrates that $F_1$ grows proportionally with $M$
for small masses, but then rapidly saturates to $F_1^{sat} \simeq
0.435$~N. $F_1(M)$ is only weakly frequency dependent, again
indicating that in the contact regime a well-defined quasi-static
regime is probed. Note that for all $f$ a small overshoot occurs for
intermediate values of $M \approx 100$~g.

This behavior is reminiscent of the Janssen effect for which the DC
bottom force $F_b^{dc}$ goes as \cite{eric,janssen,rev_degennes}:
\begin{equation}\label{janssendc}
F_b^{dc} = F_{sat}^{dc}\left[1 - \exp\left(-M g /
F_{sat}^{dc}\right)\right],
\end{equation}
where $F_{sat}^{dc}$ is the saturation force. Figure~\ref{jansen}(b)
shows, however, that the AC bottom force $F_1$ significantly
deviates from the static Janssen result.  We conclude that, even in
the limit of weak vibrations, the force variation $F_b^{ac}$ is {\em
not} simply related to the steady force $F_b^{dc}$.

\begin{figure}[tbp]
\includegraphics[width=\columnwidth]{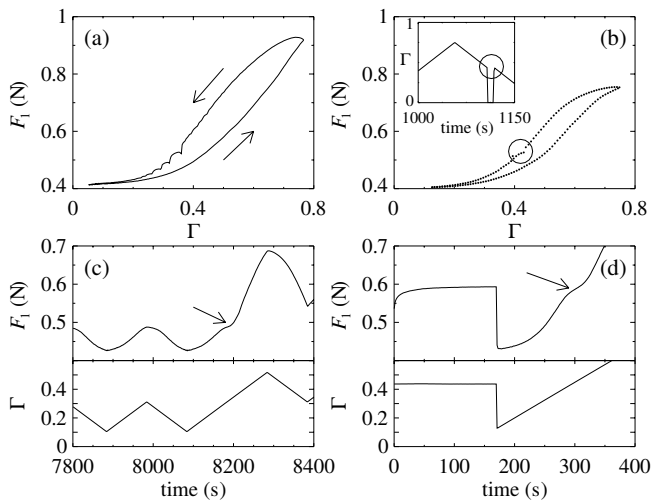}
\caption[]{Memory effects ($M\!=\!200$~g and $f\!=\!80$~Hz). (a)
Hysteresis loop. (b) Hysteresis persists when $\Gamma$ is suddenly
set to zero and then rapidly ramped back up, indicated by the circle
(see inset for details of ramp). (c-f) Two examples of subtle memory
effects.}\label{fig_memory} \vspace{-4.1mm}\end{figure}

{\em Hysteresis and Memory} --- So far, our data is consistent with
a simple nonlinear response of the material, but
Fig.~\ref{fig_memory}(a) indicates that when $\Gamma$ is ramped up
and down in the nonlinear regime, $F_b^{ac}$ is also hysteretic. The
magnitude of hysteresis is only weakly dependent on the driving
frequency and increases with $M$ and $\Gamma$ similar to the
magnitude of overall nonlinearity. The hysteresis is nearly
independent of the sweep speed (for sweep durations longer than
$\approx 100/f$), and the force configuration thus depends on the
driving history. Fig.~\ref{fig_memory}(b) illustrates that these
configurations can be ``frozen'', since the system returns to the
upper branch of the hysteresis loop after the driving is switched
off and then rapidly ramped up again. The system exhibits additional
subtle memory effects: after a fully annealed system is subject to a
number of small amplitude sweeps in $\Gamma$, and then $\Gamma$ is
ramped beyond the peak value of the small sweeps, a clear ``kink''
in the $F_1$ curve is exhibited [Fig.~\ref{fig_memory}(c,d)]. When a
fully annealed system is driven at a fixed $\Gamma$, it
``remembers'' this value when $\Gamma$ is rapidly decreased and then
ramped past the initial fixed value [Fig.~\ref{fig_memory}(e,f)].

{\em Discussion} --- Our experiments exhibit rich dynamical behavior
of weakly excited ($\Gamma<1$) granular media which can be dominated
by either grain motion or by contact force variations. That grain
motion and compaction occur in loose samples vibrated at low
frequencies and $\Gamma<1$ is maybe not surprising, although we are
unaware of systematic studies of compaction in this regime
\cite{nmat}. In compacted samples, both grain dominated (impact) and
force dominated (contact) regimes can be distinguished with $f$
playing a crucial role. For low $f$, the system is in the contact
regime, and we observe little systematic variations with $f$. We
therefore interpret the nonlinearities, memory effects, and
hysteresis of the forces in response to variations in the driving
strength as quasistatic, and not related to the excitation of sound
waves. We have not identified any theoretical or numerical
descriptions of these surprisingly strong effects. Exploratory
experiments in various columns with rough walls and for particles of
different sizes further illustrate the robustness of these phenomena
\cite{in_preparation}.

How should weakly vibrated granular systems be viewed? The contact
laws between elastic bodies display a variety of nonlinear and
hysteretic behaviors \cite{kljohnson,hhprl}. A weakly driven
granular assembly apparently amplifies the local nonlinearities
present in the hertzian contacts and the friction law. We propose
that force networks ``activated'' by weak vibrations explore many
different configurations consistent with the overall boundary
conditions for the stress \cite{jacco,bouchaud}. Such activated
force networks could possibly play a role in creep flows, which
occur far away from shear zones, and more generally in any granular
system in which tiny relative grain motions are excited. In this
sense, weakly driven granulates cannot be thought of as ordinary
solids.

The authors thank CATS and the Netherlands Organization for
Scientific Research (NWO) who supported numerous visits to NWU
during which this work was carried out.


\begin{thebibliography}{99}
\bibitem{thd} The harmonic distortion of $F_b^{ac}$ is defined as
$\sqrt{\Sigma_{i=2}^\infty F_i^2}/F_1$.
\end{thebibliography}
\end{document}